\documentclass[twocolumn,aps,prl,epsf]{revtex4}
\usepackage{amsmath}
\usepackage{epsfig}
\usepackage{epsf}
\usepackage{array}

\begin{document}

\title{Electron-lattice coupling, orbital stability and the phase 
diagram of Ca$_{2-x}$Sr$_x$RuO$_4$}
\author{Satoshi Okamoto and Andrew J. Millis}
\affiliation{Department of Physics, Columbia University, 
538 West 120th Street, New York, NY 10027}
\date{\today}

\begin{abstract}
Hartree-Fock calculations are presented of a theoretical model describing the 
Sr/CaRuO$_4$ family of compounds. Both commensurate and incommensurate magnetic
states are considered, along with orbital ordering and the effect of lattice
distortions. 
For reasonable parameter values, interactions disfavor orbital disproportionation. 
A coherent description of the observed phase diagram is obtained.
\end{abstract}

\maketitle


The two dimensional ruthenate family Ca$_{2-x}$Sr$_x$RuO$_4$ presents a rich phase diagram
\cite{Nakatsuji00}, evolving as $x$ is varied from the Fermi-liquid-triplet superconductor 
Sr$_2$RuO$_4$ to the antiferromagnetic (Mott) insulator Ca$_2$RuO$_4$. 
The origin and indeed the nature of some of the phases remains the subject of controversy. 
Further interest has been added by surface studies \cite{Moore04}, 
which indicate that 
(contrary to intuition) the Mott state is less stable on the surface than in the bulk. 
Bulk Ca$_{1.9}$Sr$_{0.1}$RuO$_4$ exhibits a metal to insulator transition as 
the temperature is decreased below 150~K \cite{Nakatsuji00}, 
whereas surface sensitive probes place the transition at $\approx$125~K \cite{Moore04}. 
This unexpected result challenges the conventional notion that correlation effects are stronger 
at surface, and requires explanation.

In this paper, we present calculations which shed new light on 
Ca$_{2-x}$Sr$_x$RuO$_4$. While a prior theoretical literature exists
\cite{Nomura00,Fang01,Hotta02,Anisimov02,Liebsch03}, 
our analysis involves new features including study of 
heretofore unexplored parameter ranges, of incommensurate magnetic phases, 
and of coupling to the apical oxygen. 
We have also uncovered a previously unnoticed phenomenon: 
for wide and physically reasonable parameter regimes, interactions stabilize the system 
against orbital disproportionation. 

We study a model Hamiltonian derived from a tight-binding approximation to the calculated
band structure \cite{Oguchi95} and supplemented by electron-electron and electron-lattice
interactions:
$H=H_{band}+H_{e-e}+H_{e-latt}+H_{latt}$.  The near-Fermi-level states 
are derived from 
Ru $t_{2g}$ symmetry $d$-states (admixed with oxygen), well described by a tight-binding 
dispersion 
$H_{band} = H_{xy} + H_{xz,yz}$ with 
\begin{eqnarray}
H_{xy} &=& \sum_{k \sigma} \varepsilon_k^{xy} d_{k \sigma xy}^\dag d_{k \sigma xy}, \\
H_{xz,yz} &=& \sum_{k \sigma} (d_{k \sigma xz}^\dag d_{k \sigma yz}^\dag) \!\!
\left( 
\begin{array}{cc} 
\varepsilon_k^{xz} &  \varepsilon'_k \\ 
\varepsilon'_k & \varepsilon_k^{yz} \\ 
\end{array} 
\right) \!\!\!
\left( 
\begin{array}{c} 
d_{k \sigma xz} \\ 
d_{k \sigma yz} \\ 
\end{array} 
\right) ,
\label{eq:Hband}
\end{eqnarray}
where $\varepsilon_k^{xy} = 
\varepsilon_0^{xy} -2t (\cos k_x + \cos k_y) - 4 t_1 \cos k_x \cos k_y$, 
$\varepsilon_k^{xz} = \varepsilon_0^{xz,yz} -2t_2 \cos k_x - 2t_3 \cos k_y$, 
$\varepsilon'_k = -4t_4 \sin k_x \sin k_y$ and 
$\varepsilon_k^{yz} (k_x,k_y) = \varepsilon_k^{xz} (k_y,k_x)$. 
Quantum oscillation measurements imply 
as $\varepsilon_0^{xy}=0.62, \varepsilon_0^{xz,yz}=0.32, 
t=0.42, t_1=0.17, t_2=0.30, t_3=0.03$ and $t_4=0.04$~(eV) \cite{Bergemann00}.
These band parameters give $n_{xy}=1.31$ and $n_{xz}=n_{yz}=1.35$, almost $4/3$. 
A crucial question, discussed in more detail below, is how the orbital occupancies change 
as parameters are varied.
The electron-electron term, $H_{e-e} = \sum_i H_{e-e}^{(i)}$ is 
\begin{eqnarray}
H_{e-e}^{(i)}&=&U \sum_a n_{i a \uparrow} n_{i a \downarrow} 
+(U'-J) \sum_{a > b, \sigma} n_{i a \sigma} n_{i b \sigma} \nonumber \\
&&+U' \sum_{a \ne b} n_{i a \uparrow} n_{i b \downarrow} 
+J \sum_{a \ne b} d_{i a \uparrow}^\dag d_{i b \uparrow} 
d_{i b \downarrow}^\dag d_{i a \downarrow}.
\label{eq:Hee}
\end{eqnarray} 
We have written in Eq.~(\ref{eq:Hee}) in the 
conventional notation \cite{Mizokawa95} in which 
$U$, $U'$, $J$,  are the intraorbital Coulomb, interorbital Coulomb, 
interorbital exchange,  interactions, respectively and we have omitted
a pair-transfer ($J'$) interaction which does not affect our results.
For $d$-orbitals, $U$, $U'$, and $J$ are functions of only 
two of independent linear combinations of the fundamental atomic physics (Slater) parameters 
$F_0,F_2, F_4$. One combination involves the multiplet averaged interaction $F_0$, along 
with a small admixture of $F_2$ and $F_4$
and is expected to be strongly renormalized by solid state effects \cite{Marel88}; 
the other involves the combination of $F_2$ and $F_4$ which determines $J$ and is expected 
to be less affected by solid state effects. The precise linear combinations depend on the 
strength of the ligand field and are not important here, but in cubic symmetry 
$U=U'+2J$. 
(In a tetragonal environment small corrections occur; 
neglected here for simplicity.)
We therefore set $U=U'+2J$  and explore a range of 
$U$ and $J$ values, which is equivalent to changing $F_0$ and bandwidth.  
Finally, the electron-lattice coupling and lattice distortion energy are given by 
\begin{eqnarray}
H_{e-latt} + H_{latt}= \lambda \sum_i \tau_{3i} Q_{i} + \frac{1}{2} K \sum_i Q_i^2.  
\label{eq:Hlatt}
\end{eqnarray}
Here, $\tau_{3} = n_{xy} - \frac{1}{2}(n_{xz} + n_{yz})$, and $Q$ represents 
a normal coordinate of RuO$_6$ distortion (apical oxygen displacement). 
$\lambda$ and $K$ are the electron-lattice coupling constant and the spring constant, 
respectively. 

We use the Hartree-Fock approximation to determine the ground state phase diagram by 
comparing energies of different phases including paramagnetic (PM), 
a ferromagnetic (FM) state, 
commensurate antiferromagnetic states with $\vec q_C =(\pi, 0)$ (C-AFM) 
and $\vec q_G =(\pi, \pi)$ (G-AFM), and 
an incommensurate magnetic (ICM) state with 
$\vec q_{IC} =(2 \pi/3, 2 \pi/3)$. 
$\vec q_{IC}$ is close to the momentum $\vec q^* \simeq (0.69 \pi, 0.69 \pi)$ 
where the susceptibility $\chi$ has a local maximum 
due to the near nesting of quasi-one-dimensional $\{xz, yz\}$-bands. 
Strong magnetic scattering peaked near $\vec q_{IC}$ 
has been experimentally observed in Sr$_2$RuO$_4$ \cite{Sidis99}. 
We also considered the orbital ordering (OO), i. e., changing the relative 
occupancy of $xy$, $xz$ and $yz$ states. Uniform states were found,
but two-sublattice states such as to 
those reported in Ref.~\cite{Hotta02} were never found to minimize the energy 
for the parameters studied. 

Before presenting our results we discuss a simple one-atom calculation.
In conventional atomic physics, one minimizes the energy using a state 
which corresponds to a definite electron
occupancy, and finds (for $n=4$) that the ground state corresponds to maximal spin and 
orbital angular momentum, with  excited state energies 
determined by the higher Slater parameters, 
$F_2$, $F_4$ i. e., by $J$. For a solid state environment, it is more appropriate to 
minimize using a density matrix corresponding to the desired mean 
charge density per orbital $n_a$ and spin density per orbital $m_a$. 
In this case
the $F_0$ (i. e., $U$) parameter also contributes to energy differences. 
It is convenient to express the densities in terms of $n_0=n_{xy}+n_{xz}+n_{yz}$, 
$\tau_3$ [defined below Eq.~(\ref{eq:Hlatt})] and 
$\tau_2 = \frac{\sqrt{3}}{2}(n_{xz}-n_{yz})$ and similarly for $m$. 
For paramagnetic states ($m_a=0$) and fully polarized states ($m_0=2$), we find 
\begin{eqnarray}
\hspace{-2.3em}&&E_{int}^{para}(n_0=4)=-\frac{1}{6}(U-5J)(\tau_3^2+\tau_2^2), \\
\hspace{-2.3em}&&E_{int}^{polarized}(n_0=2,m_0=2)=-\frac{1}{3}(U-3J)(m_3^2+m_2^2). 
\end{eqnarray}
Thus, in a solid  state environment, 
interactions may either promote or inhibit orbital disproportionations; in
the present Hartree-Fock approximation 
for paramagnetic (fully polarized) states
disproportionation is suppressed for $J>U/5$ ( $J>U/3$). 

\begin{figure} [t]
\begin{center}
\includegraphics[scale=0.5,clip]{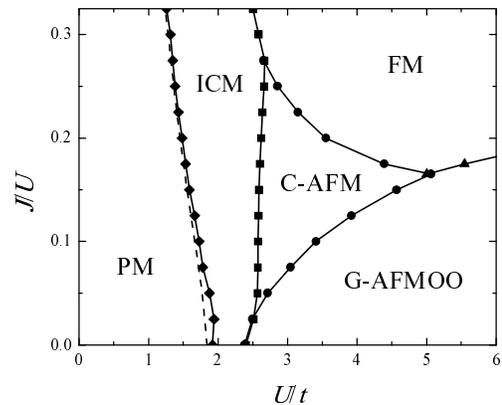}
\end{center}
\caption{Phase diagram without electron-phonon coupling for band 
parameters corresponding to 
$n_{xy}=n_{xz}=n_{yz}=4/3$ at $U=0$.  
PM: paramagnetic state, 
FM: ferromagnetic state, 
C-AFM: C-type antiferromagnetic state [$\vec q_C=(\pi,0)$],
G-AFM: G-type antiferromagnetic state [$\vec q_G=(\pi,\pi)$],
ICM: incommensurate magnetic state [$\vec q_{IC}=(\frac{2\pi}{3},\frac{2\pi}{3})$].
OO: orbital disproportionation with $n_{xy} \neq n_{xz,yz}$. 
Only the G-AFM state is insulating. 
Symbols connected by solid lines indicate phase boundaries 
(first order, except for the ICM/PM boundary)  between the indicated
states. Broken line: second-order phase transition from PM to ICM 
with $\vec q^*$ 
determined by the susceptibility maximum.
As $U/t \rightarrow \infty$ the G-AFMOO-FM phase boundary asymptotes to $J/U=1/3$ 
as discussed in the text.}
\label{fig:PD_uj}
\end{figure}

Figure~\ref{fig:PD_uj} shows the numerically obtained ground-state phase 
diagram 
without electron-phonon coupling as a function of $U/t$ and $J/U$. 
At large $U$, two phases are observed; a ferromagnetic metal
with a small degree of orbital
disproportionation (FM) and an insulating
antiferromagnetic phase  (G-AFMOO).
In the G-AFMOO state, the $xy$-orbital is occupied by two electrons 
and the other two electrons sit on the $\{xz, yz\}$-orbitals forming a half-filled band 
gapped by the AF order. 
Even in the absence of long range order, the strong correlations and commensurate filling 
of the $\{xz, yz\}$-bands would lead to Mott insulating behavior. This is the state 
proposed by Nakatsuji and Maeno \cite{Nakatsuji00} for Ca$_2$RuO$_4$. 
Because it involves orbital disproportionation
the G-AF phase is only stable for small $J/U$. 
Turning now to the small $U$ regime, we find that as $U/t$ is 
increased the first phase transition
is of second order, and is to an ICM state 
characterized by a wave vector $q^* \simeq (0.69\pi,0.69\pi)$
determined by near nesting of the  
$\{xz,yz \}$ bands. $\chi$ also has a strong peak
at a much smaller $q_2 \sim (0.5,0.5)$, arising from the $xy$ band,
but in the present calculation the $q^*$ instability is slightly favored. 
We suggest that the IC phase is  the one observed in the actual materials for
$0.2<x<0.5$. As $U$ is further increased a first order transition occurs to a phase which 
is FM or C-AFM depending on $J/U$. 
Note that in computing the energy of the ICM phase we approximated the ICM vector by 
$\vec q_{IC}=(2 \pi/3, 2 \pi/3)$. 
The errors due to this approximation may be estimated from the difference between 
broken and solid lines in Fig.~\ref{fig:PD_uj}.

We now interpret experiment in terms of our results,
beginning with Sr$_2$RuO$_4$. Quantum oscillation measurements 
(Table~6 in Ref.~\cite{Bergemann00}) directly determine both spin polarization and 
quasiparticle mass enhancement and reveal a Stoner factor (enhancement of $\chi$ above that
due to the quasiparticle density of states) of $4-5$. 
Within our approximation this corresponds to $2<U/t<2.3$ (weakly dependent on $J/U$), 
i. e., to a $U/t \approx 0.8$ of the critical value for a ferromagnetic instability 
(preempted here by the ICM state). 
We also find that the three dimensional model has a second order PM-FM transition 
at $U+2J \simeq 2.5t$ and for $U+2J \simeq 3t$, $m \sim 1 \mu_B$
consistent with observation \cite{Longo68}.
Thus we argue that a moderate interaction, 
e. g., $U \simeq 2t$ and $J \simeq 0.6t$, 
provides a good description of the Sr-ruthenates. 

Our estimate of the interaction strength implies that a $\sim 10~\%$ decrease in $t$ 
(a change of the magnitude expected when Sr is replaced by Ca) will drive
a transition into an ordered state. 
Over a wide range of the phase diagram, our calculation indicates that the first 
instability is to an ICM state driven by near Fermi surface nesting. Indeed we find that 
this instability occurs at $U/t<2$, i. e., where the Stoner factor is
less than 4. However, beyond-Hartree-Fock-corrections are expected to shift the ICM phase 
boundary substantially, both because we are dealing with low dimensional magnetism, 
on which quantum fluctuations have substantial effect, 
and because inelastic scattering will weaken the susceptibility peaks. 
Further, the two possible ICM wave-vectors lead to very similar critical $U$'s; 
beyond Hartree-Fock effects are expected to control whether 
$\{ xz,yz\}$-band dominated $\vec q^* \simeq (2\pi/3,2\pi/3)$ or 
$xy$-band dominated $\vec q_2 \simeq (0.5,0.5)$ magnetism occurs. 
Thus, while it is natural to expect a second order magnetic transition to occur
as Ca is substituted for Sr, the specifics
depends on details beyond the scope of our calculation.
We argue that the most natural interpretation of the data is that the transition observed
at $x=0.5$ is the PM-ICM transition we found, but that 
the relatively larger fluctuation corrections to the ICM state shift the ICM critical point 
close to the FM critical point (Stoner factor at $x=0.5$ $\approx 25$). 
In the actual materials, the $T=0$ phase boundary to the 
ICM state is close to the onset of
a lattice distortion \cite{Friedt01} which 
distinguishes the [1,1] and [1,$\bar 1$] directions of our 2$d$ lattice. 
In the distorted phase the two ICM vectors $\vec q_{IC}$ and 
$\vec q_{\overline {IC}} =(q,-q)$ become inequivalent, so 
diagonal ICM order and the lattice distortion couple. 
A minimal description is 
\begin{eqnarray}
{\cal F} = {\cal F}_S (S_{IC}^2, S_{\overline {IC}}^2) 
+\gamma (S_{IC}^2 - S_{\overline {IC}}^2) X + \frac{1}{2} C X^2, 
\label{landau}
\end{eqnarray}
where $S_{IC(\overline{IC})}$ is the magnetic order parameter associated with 
$\vec q_{IC} (\vec q_{\overline {IC}})$,
$X$ denotes the distortion amplitude at $0.2<x<0.5$ and
$\gamma$ and $C$ are a coupling constant and an elastic constant, respectively. 
Eq.~(\ref{landau}) shows that the magnetic transition will  induce the
lattice transition, consistent with the observed coincidence of the two
transitions at $x=0.5$, $T=0$. Alternatively, a primary lattice transition could induce 
a first order magnetic transition. In either case
the observed \cite{Nakatsuji00} difference in transition temperatures  follows from the 
strong effect of fluctuations of low-dimensional magnetism. 
Note, however, diagonal $(q,q)$ wave vectors are always found to be favored over 
``face'' $(q,0)$ vectors. Bragg peaks corresponding to the 
ICM phase have not been observed. 
Friedt {\it et al}. do observe low-energy incommensurate fluctuations at 
$\vec q \simeq (0.6,0)$ \cite{Friedt03} 
but in our calculations $\chi$ is always greater at $(0.5,0.5)$.
The coincidence of magnetic and structural transitions also favors diagonal order. 

\begin{figure}[b]
\begin{center}
\includegraphics[scale=0.5,clip]{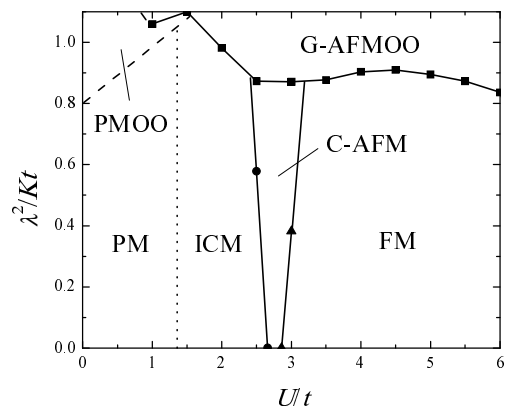}
\end{center}
\caption{Phase diagram as a function of $U/t$ and $\lambda^2/Kt$ for $J/U=0.25$ 
and $n_{xy}=n_{xz,yz}=4/3$.
Solid lines: first-order phase 
transitions. Dotted line: second-order transition from PM to ICM. 
Broken line: transition to quadrupole ordering with $\tau_3 > 0$; 
weakly first-order because cubic terms 
are allowed in the free energy. 
}
\label{fig:PD_uph}
\end{figure}

Further decrease of $t$ leads to a first order transition into a strongly magnetic state. 
With the Hartree-Fock approximation, the first transition is to either the C-AFM phase or
the FM, depending on the value of $J/U$. As far as we know, neither of these phases is 
observed experimentally. To account for the direct transition into the G-AFM phase which 
is apparently observed, 
an additional coupling which favors orbital disproportionation must be included. 
A very natural possibility is coupling to a ``$Q_3$-type'' lattice distortion 
(volume-preserving increase in Ru-apical-oxygen distance and decrease in Ru-planer-oxygen 
distance). 
Fig.~\ref{fig:PD_uph} shows results obtained using Eq.~(\ref{eq:Hlatt}), for $J/U=0.25$. 
We see that the range of G-AFM phase is increased, and for strong enough electron-phonon
coupling, a direct transition from ICM to G-AFM is possible. Also within this model, 
a lattice distortion and a hardening of the phonon frequency occur at the transition to
the G-AFM phase, again in at least qualitative agreement with the experiment\cite{Rho03}.
Pressure experiments on Ca$_2$RuO$_4$ observe a transformation to a metallic phase with a
small FM moment.
Pressure increases $t$ and $K$, corresponding to diagonal motion in Fig.~\ref{fig:PD_uph}. 
The FM phase shown is highly polarized in contrast to that found in Ref.~\cite{Nakamura02}.
A weakly polarized FM phase exists as a local free energy minimum in the region labeled 
ICM; it is conceivable that pressure suppresses the ICM phase, revealing the FM one 
but this is not found in our theory. 

\begin{figure}[t]
\begin{center}
\includegraphics[scale=0.5,clip]{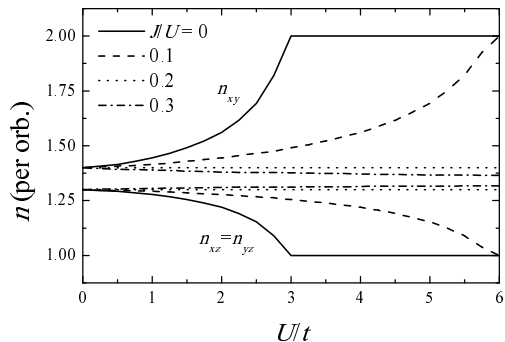}
\end{center}
\caption{Orbital occupancies in PM phase as functions of $U/t$ for a band structure 
$n_{xy}=1.4$ and $n_{xz}=n_{yz}=1.3$ at $U=0$. 
}
\label{fig:n_u}
\end{figure}

The results presented so far are obtained for a band structure corresponding 
to $n_{xy}=n_{xz}=n_{yz}=4/3$ at $U=0$. Fig.~\ref{fig:n_u} shows the evolution of orbital 
occupancies in the PM phase for different $J/U$ and band structure such that at $U=0$ 
$n_{xy}=1.4$, $n_{xz,yz}=1.3$. We see that, as expected from the simple estimate, 
the orbital disproportionation is strongly enhanced by correlation effects if $J$ is less
than critical value $J_c$ ($=0.2U$ in the Hartree-Fock approximation used here) and 
is suppressed if $J$ is greater than the critical value. 
The strong coupling between different types of order and the orbital occupancy means 
that for $J/U<0.2$ the phase diagram is very sensitive to the $xy/\{xz,yz\}$ level 
splitting, but for $J/U>0.2$ is much less so. 

Our results suggest a possible interpretation of the puzzling surface experiments by
Moore {\it et al}. who find that the surface of Ca$_{1.9}$Sr$_{0.1}$RuO$_4$ remains metallic down to a lower temperature (125K) than does the bulk material (150K) and that
both on the surface and in bulk the metal-insulator transition
is accompanied by a $Q_3$-type distortion of magnitude $\sim0.02$. 
The surface phonon frequency of the apical oxygen mode is 10~\% larger than bulk, 
implying a 20~\% increase in $K$ and thus a 20~\% decrease in $\lambda^2/Kt$. 
As seen from Fig.~\ref{fig:PD_uph}, a change of this magnitude makes the insulating phase
substantially more difficult to access. A full theory requires a study of 
the $T$ dependence, for which Hartree-Fock is unreliable. 
Other issues also remain unresolved, including the roles played by the frozen-in $Q_3$ and 
$Q_0$ distortions occurring at the surface and the additional lattice strains caused by 
the bulk transition. Further investigation using dynamical-mean-field methods and 
a detailed treatment of the lattice effects would be desirable. 

Our approach is most similar to that of
Nomura and Yamada \cite{Nomura00}, who performed Hartree-Fock studies of the model 
studied here, but with band parameters corresponding to slightly 
broader 1$d$ bands. They assumed $U'/U=0.5$, $J/U=0.25$, (`orbitally stable'), 
considered only G-AFM and FM ordered phases and did not include electron-lattice 
coupling, but did investigate effects of changes in relative level energies.
In a closely related work, Fang and Terakura used LSDA band theory methods to investigate 
FM and G-AFM states \cite{Fang01}. 
Their results, in particular their prediction of a direct PM-FM transition
as bandwidth is reduced, correspond reasonably well to our results 
with $J/U \approx 0.3$, the ICM phase omitted. 
Very recently, the calculation was refined to include a LDA+$U$ treatment of 
Ca$_2$RuO$_4$ and the experimental structure, 
and results for optical absorption were presented. 
Hotta and Dagotto \cite{Hotta02} studied the model via a combination of mean-field results 
and numerical studies of small systems and showed 
the importance of coupling to the shape changes of the RuO$_6$ octahedron. 
However, their phase diagrams feature phases with complicated spatial 
structures which we are unable to stabilize. It is possible that the spatial structures 
found in Ref.~\cite{Hotta02} are due to boundary effects in small size systems 
accessible numerically. 
Anisimov and co-workers \cite{Anisimov02} used ``dynamical-mean-field'' and LDA+$U$ method 
to study the $U'=J=0$ model. 
At strong correlation, they found an insulating ``(2,2)'' phase which 
is essentially the same as the G-AFMOO phase we find.
At slightly lesser correlation strengths, Anisimov {\it et al}. find an interesting 
``(3,1)'' phase in which 3 electrons are in the $\{xz,yz\}$ bands in 
a Mott insulating state, while the $xy$ band remains metallic. 
This work remains controversial\cite{Liebsch03,Koga04}. 
The Hartree-Fock analogue (AF state with $n_{xy}=1$) is not found here: at $J/U>0.2$ 
it is presumably disfavored by the orbital stability effects we have discussed; 
for smaller $J$ it appears to be preempted by the C-AFM and ICM phases (not considered in
Refs.~\cite{Anisimov02,Liebsch03,Koga04}). 
Very recent ARPES experiments indicate $n_{xz}=n_{yz}=4/3$ \cite{Wang04}, 
inconsistent with the (3,1) phase, but consistent with our orbital stability arguments of 
$J/U > 0.2$. 

To summarize: we have shown that moderate interactions, a reasonable
$J/U$ and electron-lattice interactions
can account for the Ca$_{2-x}$Sr$_x$RuO$_4$ phase diagram, and argued. 
We argued that the materials are in the regime in which 
disproportionation is suppressed by interactions. 
Determining the susceptibility to orbital 
disproportionation  in other compounds and other 
multiplet structures is an important open issue. Investigation of 
ICM/lattice-distortion instability ($x\sim0.5$) via Landau theory methods and study, using
dynamical-mean-field methods, of the G-AFM boundary are important directions. 

We thank J. Zhang, H. Ding and E. W. Plummer for stimulating discussions and 
sharing unpublished data.
This research was supported by NSF DMR-0338376 (A.J.M.) and 
JSPS (S.O.).


\end{document}